# On the formation process of silicon carbide nanophases via hydrogenated thermally induced templated synthesis**

*Mark. H. Rümmeli[1*], Ewa Borowiak-Palen[1,2], Thomas Gemming[1], Martin Knupfer[1], Kati Biedermann[1], Ryszard J. Kalenczuk[2], Thomas Pichler[1]*

1 Leibniz-Institute for Solid State and Materials Research Dresden, P. O. Box 270016, Dresden D-01171, Germany

2 Institute of Chemical and Environment Engineering, Technical University of Szczecin, Poland

* m.ruemmeli@ifw-dresden.de

** This work is partially supported by DFG PI 440/1 & MHR thanks the EU for support through an MC Fellowship. We are grateful to K. Müller, R. Hübel and S. Leger for technical assistance and O. Jost for delivering the sample of SWCNT.

The massive and ever growing interest in carbon nanotubes (CNT) has and is stimulating the synthesis of various hetero-CNT. The mechanical, chemical, thermal and electronic properties (wide band gap) of SiC make it an attractive material to include in the family of nanotubes. Indeed, there are great hopes for SiC nanostructures in applications such as nanosensors and nanodevices operable at high temperature, high frequency and high power [1, 2]. Another potential use is as a support material in the catalysis field[3]. However, the production of SiC nanotubes remains a challenge, not least due to the numerous polytypes for SiC. These polytypes are orientational sequences by which the layers of tetrahedrally bonded C and Si atoms are stacked. None-the-less, the synthesis of tubular SiC structures with nanometer scaling has been shown to be possible through two techniques; shape memory synthesis[3-5], and template synthesis[6]. In this paper we present a thermally induced templated synthesis in which a reaction of SWCNT at high temperature in the presence of Si vapour in a reduced carrier gas atmosphere leads to a variety of SiC nanostructures depending on the conditions. Outside of information on the actual nanostructures themselves, the data obtained helps elucidate what reaction processes take place, which is also of relevance to techniques where CNT/Si contacts are formed by annealing CNT on a Si substrate e.g. when forming a Schottky diode by placing a SiC junction on metallic SWCNT[7].

Various samples where prepared with reaction times ranging from 0.5 – 20 h. All samples yielded soot containing two species; a grey material in the center and the remainder was black concomitant with the starting SWCNT (fig. 1a). Here we discuss only the analysis of the grey product (unless otherwise stated). The amount of grey material increased with reaction time and many changes are observed in the produced nanostructures depending on the reaction time, some of which, are observable from their surface morphology and others not. The shortest reaction time of 0.5 h yielded SiC nanorods with a C cladding (fig. 1b), further increasing the reaction time yielded pure SiC nanorods (fig. 1c) that then transform into SiC nanotubes, then SiC nanocrytals, which, with yet longer reaction times disappear altogether. This rather striking observation is indicative of some decomposition process. These various nanostructures were only observed when using $NH_3$ as the carrier gas. Using $N_2$ as the carrier gas yielded only SiC nanorods which where narrower in diameter than their counterparts produced with $NH_3$, suggesting H plays a key role in the formation of the transformed nanostructures.

Bulk stochoimetric studies of the samples began with electron diffraction studies that showed all samples to be a mix of SiC α and β phases with random orientation (fig 1d). The SiC Raman response at around 900 cm$^{-1}$ (folded longitudinal optical mode) showed a sharp shoulder on the higher frequency side and the low frequency side decreased slowly (inset fig. 1d) which is an indication that the α phase consists of many polytypes [8].

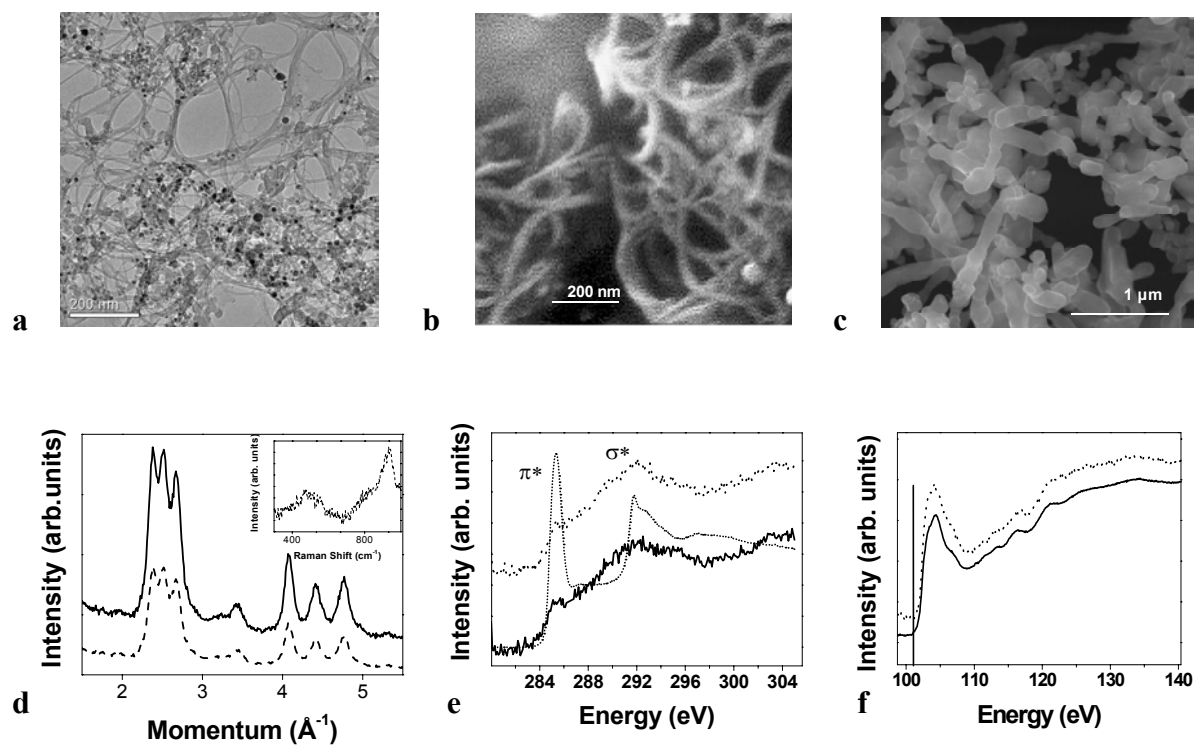

**Figure 1**. a. TEM of SWCNT. b. TEM of SiC nanorods with C cladding. c. SEM of SiC nanorods/nanotubes. d. Electron diffraction of SiC nanorods/nanotubes (8 h sample) dashed line; SiC reference sample - solid line; inset: Raman spectra of SiC nanorods/nanotubes (8 h sample). e. EELS spectra of C1s core edges: SiC reference - solid line, SiC nanorods/nanotubes (8 h sample) – dashed line, SWCNT - dotted line. f. Si2p core edges: SiC reference - solid line, SiC nanorods/nanotubes (8 h sample) – dashed line, vertical line indicates edge postion for crystalline Si. (N.B. Graphs are offset for convinience)

High-resolution EELS studies confirmed the samples altered bonding environment which is easily seen by the reduction of the π* resonance of the C edge for a typical SiC sample and a SiC reference as compared to the starting SWCNT (fig. 1e). In addition the Si edges show no superposition of spectral features from SiC and Si [9]. In general, the Si peaks from our samples increase or decrease slightly relative to that of the reference sample, pointing to slight changes in electron density at the Si site for the nanostructures (Fig. 1f).

Transmission electron microscopy (TEM), including cross-sectional EELS, allowed us to perform studies on a local scale of the produced samples. These studies not only revealed the structural nature of the SiC nanostructures but also the role of H in the reaction process, which begins with SWCNT bundles and Si powder (fig. 2 i).

Analysis of the black material showed it consisted primarily of SWCNT with small localized regions where SiC crystallization has begun and illustrates the earliest stages of the reaction process are localized (fig. 2 ii). These local crystallization sites can then grow laterally and also outward due excess C that diffuses out from crystallized regions forming a C outer layer. This stage (0.5 h) is observed as SiC nanorods with a carbon cladding with diameters ranging from 19 – 60 nm (fig. 2 iii). Eventually the C source is used up from the C cladding and the lateral SWCNT bundle, yielding a SiC nanorod (fig2. iv). One might then expect that at this stage the reaction ceases and this is indeed the case for samples prepared in $N_2$; a thermal reaction also noted by others e.g. [7, 10, 11]. However, when $NH_3$ is used, a very striking transformation process begins. At the reaction temperature used $NH_3$ decomposes to N and H, thus, it is H that is responsible for the transformation process.

The sample prepared with a 5 h reaction time contained a small fraction of nanocrystals (< 5%), and an even quantity of SiC nanorods and porous nanorods (fig. 2 iv & v respectively). In addition, the diameters of the porous nanorods (d = 40 – 100 nm) tended to be larger than those of the nanorods (d = 20 – 60 nm) and both nanorod species were on the whole larger than the C cladded nanorods. A further increment in the reaction time to 8 h yielded 3 types of species (aside from a small fraction of SiC nanocrystals); nanorods (d = 40 – 80 nm) and porous nanorods (d = 70 – 150 nm) as with the 5 h sample and now, additionally, hollow nanorods or SiC nanotubes (diameter range 100 – 250 nm) which where closed ended (fig. 2 iv, v & vi respectively). The transformation of the SiC nanorods to porous nanorods and then onto nanotubes has not previously been reported and begs the question as to what exactly is the role of H in this process? It is well known that amorphous SiC can be hydrogenated (a-SiC:h) and it has been shown that H substitutes Si[12]. That H substitutes Si atoms in this reaction is highly probable, and it may also substitute C atoms, although, H preferentially bonds to C as opposed to Si [12]. Furthermore at 1450°C C-H bonds can easily break[13]. This substitution process then in essence leads to the decomposition of SiC, which will occur at a higher rate in the center of the nanorods as this is where the concentration of H is greatest due to diffusion through faults and defects. It follows then that the decomposition process will provide a source of Si and C atoms that can diffuse outward. This we observe as the outer diameters of the SiC nanorods increasing with reaction time whilst at the same time becoming, porous and then closed ended nanotubes. Not all the outwardly diffused Si and C will reform on the surface of the SiC nanorods/nanotubes such that eventually the nanotubes disintegrate leaving only SiC crystals which we observed with the sample from a 12h a reaction and, with a sufficiently long reaction time, one obtains the total decomposition of the SiC species, which we observed with a 20 h reaction sample where virtually no material remained. Thus, the formation order when H is present in the reaction, is from SWCNT to SiC nanorods cladded in C, to SiC nanorods, to porous SiC nanorods to SiC nanotubes to eventual total decomposition. This process is illustrated in Fig. 2

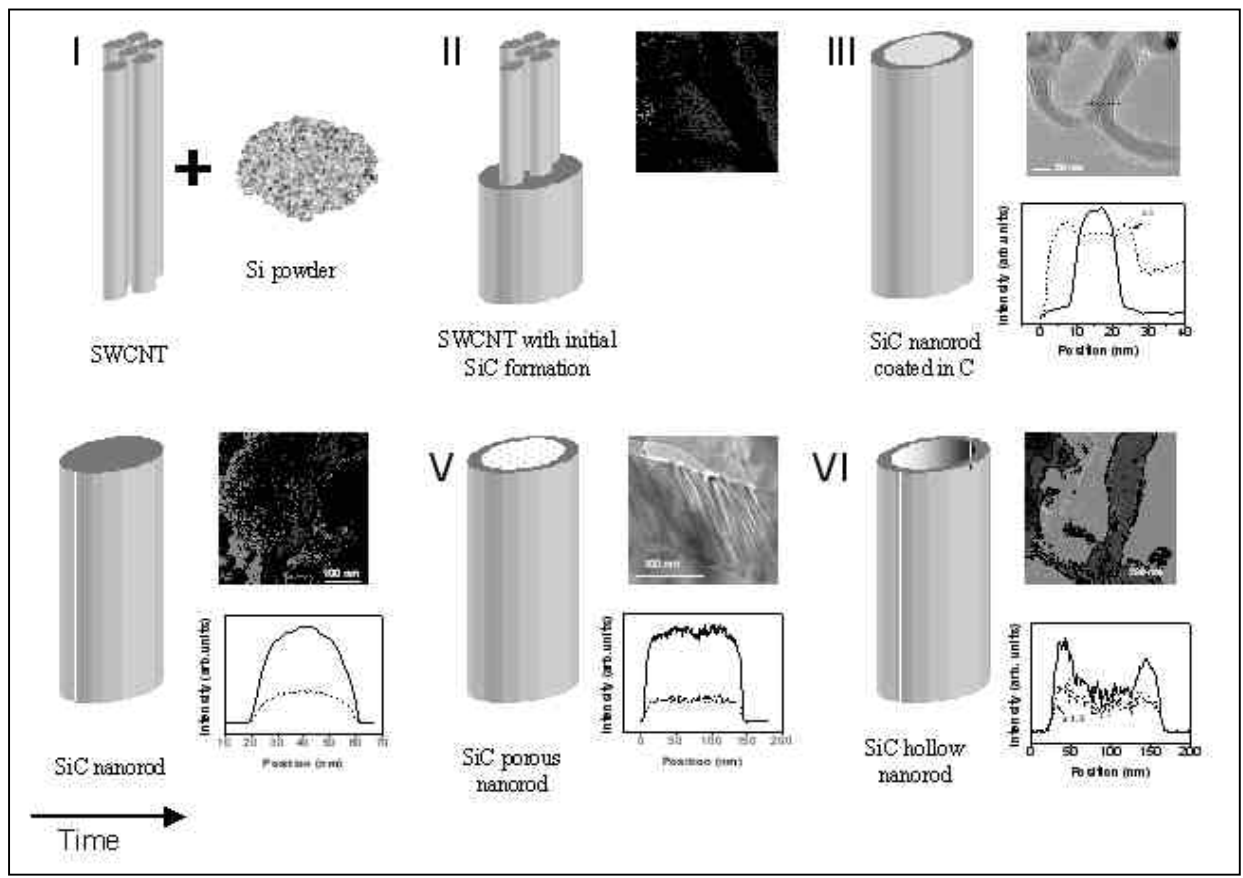

**Figure 2.** The various stages of the reaction. i. Starting materials. ii. Initial SiC formation with TEM image of the nanostructure. iii. SiC nanorod coated in C iv. SiC nanorod v. SiC porous nanorod. vi. SiC nanotube. iii to vi include a TEM image and an EELS map across the nanorod/nanotube. Reaction time increases from i through vi.

In summary, this is the first report of a reaction process that combines the substitutional ability of H on SiC and the polytypical stochiometry of SiC which provides diffusion paths for atoms that combined lead to the novel transformation of SiC nanorods to nanotubes. It is also the first report of SiC nanostructure formation via Si vapour in the presence of SWCNT. The reaction will not only help in our knowledge of hydrogenated SiC were currently very little is understood, it may also open the path, through a modification of this process, for the controlled etching of SiC films in micro/nano electronic applications.

**Experimental**

The starting material was a black powder containing 70% of SWCNT with a mean diameter of 1.25 nm produced by a laser ablation process optimized for high SWCNT yield. The carrier gas was ammonia gas, and the source of silicon was commercial silicon powder (Goodfellows, 99.99%). The SWCNT and Si were placed in an $Al_2O_3$ crucible to a ratio of 1:1. The crucible plus contents was then placed at the centre of a horizontal tube furnace (Carbolite STF 16/180). A schematic view of the experimental setup is described elsewhere[14]. The system can be evacuated allowing the base pressure to be reduced to about $10^{-7}$ mbar prior to commencement of the thermally induced templated synthesis. A highly sensitive needle valve then controls the carrier gas entry that in this instance was ammonia at a base pressure of $5*10^{-5}$ mbar. A reaction temperature of 1450°C was used throughout. Dwell times of 30 min., 4, 8 and 20 h were used with a heat up period of 30 min. and a cool down time of several hours. After the thermally induced templated synthesis the resultant material was removed from the crucible, which consisted of a portion that was light grey, and the remainder was black unreacted SWCNT. The light grey material was then carefully separated (manually) for analysis.

The morphology of the reacted nanostructures were studied using scanning electron microscopy (SEM, Hitachi S4500) and transmission electron microscopy (TEM, FEI Tecnai F30) which also allowed EELS mapping of the nanostructures. In addition, electron diffraction measurements were performed in a purpose built high-resolution EELS spectrometer described elsewhere[15]. The energy loss was set to zero. These EELS measurements probe an area of about 1 $mm^2$ and thus present information for a bulk average of the produced nanostructures. Raman measurements were conducted on a Bruker FTRaman spectrometer with a resolution of 2 $cm^{-1}$.

For the measurements tiny quantities of the product were pressed on to standard platinised microscopy grids were upon they were annealed at 450 °C for 12 h in vacuum prior to measurement so as to remove contaminants.